\documentclass[aps,showpacs,twocolumn,superscriptaddress,prl]{revtex4}
\usepackage{graphicx,color,epsfig,rotate}
\usepackage{amssymb,amsmath,bm}

\begin{document}

% ==============================================================================
%\title{Quantum Selection of Order in $XXZ$ Kagom\'{e} Antiferromagnet}
\title{Quantum Selection of Order in an $XXZ$ Antiferromagnet on a Kagom\'{e} Lattice}
% ==============================================================================

% ==============================================================================
\author{A. L. Chernyshev}
\affiliation{Department of Physics  and Astronomy, University of California, Irvine, California
92697, USA}
\author{M. E. Zhitomirsky}
\affiliation{Service de Physique Statistique, Magn\'etisme et Supraconductivit\'e,
UMR-E9001 CEA-INAC/UJF, 17 rue des Martyrs, 38054 Grenoble Cedex 9, France}
% ==============================================================================
\date{\today}

% ==============================================================================
\begin{abstract}
Selection of the ground state of the 
%nearest-neighbor 
kagom\'{e}-lattice $XXZ$ antiferromagnet   by quantum fluctuations is investigated 
by combining  non-linear  spin-wave and  real-space perturbation theories.
The two methods unanimously favor  ${\bf q}\!=\!0$ 
over $\sqrt{3}\times\!\sqrt{3}$ magnetic order in a wide range of the
anisotropy parameter $0\leq \Delta\!\alt\!0.72$.
Both approaches are also in an  accord on the magnitude of the quantum
order-by-disorder effect generated by topologically non-trivial, loop-like 
%tunneling
spin-flip processes.
A tentative $S$--$\Delta$ phase diagram of the model is proposed.
%We propose a tentative $S$--$\Delta$ phase diagram of the model and suggest
%future directions.
\end{abstract}
\pacs{75.10.Jm, 	% Quantized spin models, including quantum spin frustration
      75.30.Ds,     % Spin waves
      75.50.Ee, 	% Antiferromagnetics
      75.45.+j      % Macroscopic quantum phenomena in magnetic systems      
%      75.25.-j,     % Spin arrangements in magnetically ordered materials
%      75.10.Hk,	% Classical spin models
%      75.40.Cx	% Static properties (order parameter, static susceptibility, heat capacities, critical exponents, etc.)
%      75.40.Gb,     % Dynamic properties
%      78.70.Nx,     % Neutron inelastic scattering
}
\maketitle
% ==============================================================================

\begin{figure}[t]
\includegraphics[width=0.99 \columnwidth]{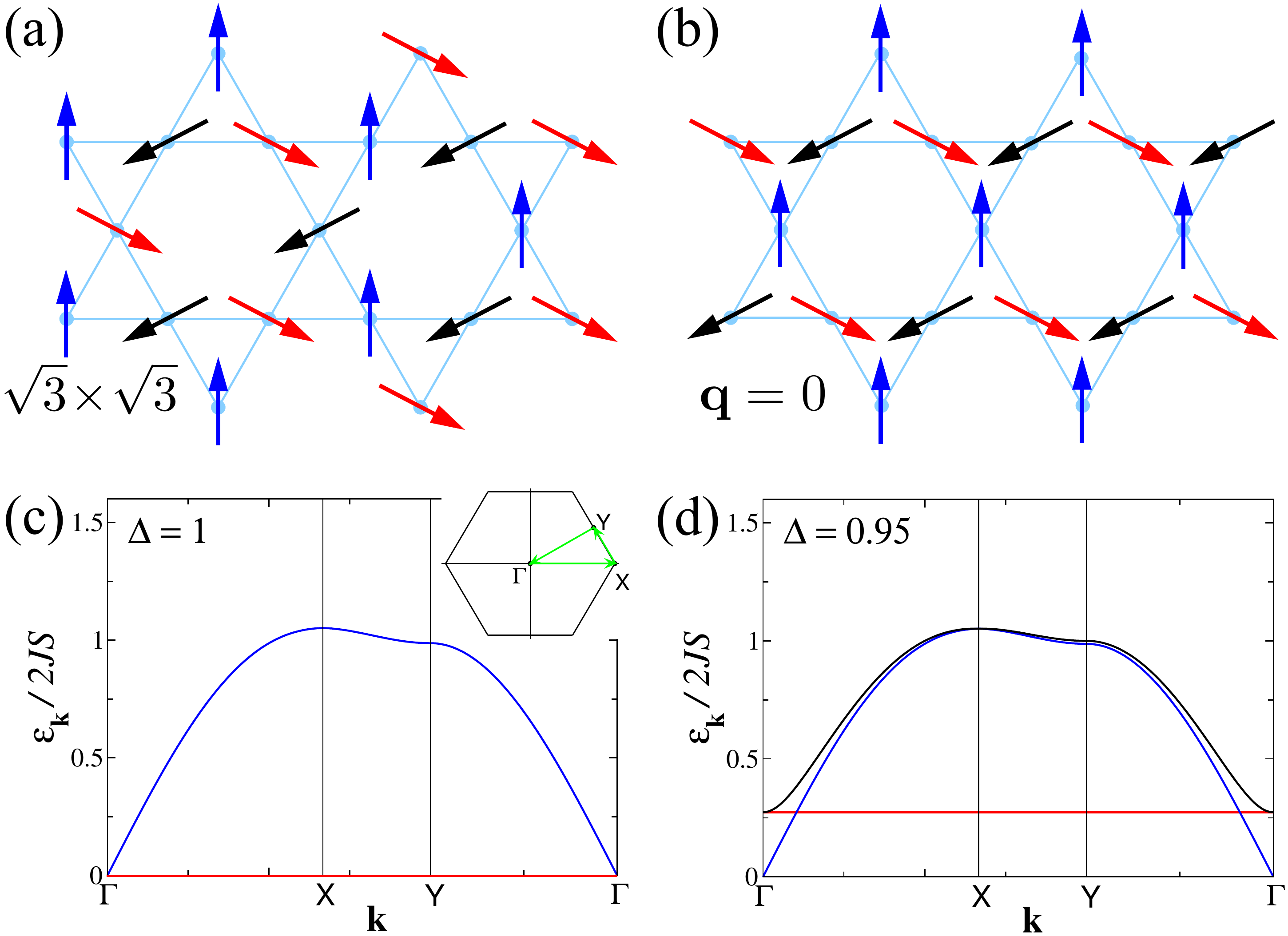}
\caption{(Color online) (a) $\sqrt{3}\times\!\sqrt{3}$ and (b) ${\bf q}\!=\!0$ spin configurations.
%Numbers label sites in the unit cell, {\boldmath{$\delta$}$_i$}  are primitive vectors of the lattice.
Dispersions of spin excitations within the harmonic approximation for (c) $\Delta\!=\!1$, and (d)
$\Delta\!=\!0.95$.
}
\vskip -0.4cm
\label{fig:states}
\end{figure}

Kagom\'{e}-lattice antiferromagnets (KGAFMs) are central to  theoretical and experimental
studies in  frustrated magnets. They host long-sought magnetically disordered spin-liquids
and intriguing valence-bond solids, exhibit order-by-disorder phenomena, and are dominated
by unconventional excitations
\cite{Zeng90,Harris92,Chalker92,Chubukov92,Sachdev92,Chalker92a,Singh92,%
Lecheminant97,Mambrini00,Nicolic03,Singh07,Matan10,Matan14,Helton07,Lee12,%
Yan11,Gotze11,Balents10,Iqbal13,Changlani14,Weichselbaum14,Nishimoto14,Rousochatzakis14,Hao10,Wan13,Taillefumier14}.
Many of these remarkable properties take their root in a massive degeneracy of the ground state
of the classical kagom\'{e} nearest-neighbor Heisenberg model.
The degeneracy can be lifted by  thermal or quantum fluctuations, or by secondary
interactions. Because of experimental realizations, order selection
by the symmetry-breaking Dzyaloshinskii-Moriya (DM) terms has been intensely studied
\cite{Zorko08,Zorko13,Yoshida12,Matan06,Yildirim06,Elhajal02,Cepas08,Messio10,Sachdev10}
and so has been the effect of further-neighbor couplings \cite{Harris92},
which lift the degeneracy within the
manifold of classical $120^\circ$ states. Two of such states, the $\sqrt{3}\times\!\sqrt{3}$
and the ${\bf q}\!=\!0$ spin patterns, are the main contenders for the ground state
from the quasiclassical perspective,\cite{Reimers93} see Figs.~\ref{fig:states}(a,b).

On the other hand, studies of  quantum  effects have been concentrated on
the Heisenberg case where most methods offer only  limited insight into \emph{how}
the ground state is selected.
In this work, we address the problem of order-by-disorder (ObD)  by quantum fluctuations
in KGAFMs using the $XXZ$ version of the nearest-neighbor, spin-$S$ model
\begin{equation}
\hat{\cal H} = J \sum_{\langle ij\rangle} \Bigl( S_i^xS_j^x + S_i^yS_j^y
+ \Delta S_i^z S_j^z \Bigr) \ ,
\label{Ha}
\end{equation}
where anisotropy is of the  easy-plane type, $0\!<\!\Delta \!<\!1$.
It is important to note that the degeneracy among the $120^\circ$ coplanar states 
of the classical $XXZ$ KGAFM
remains \emph{the same} as in the Heisenberg limit, $\Delta\!=\!1$.
Therefore, by extending the parameter space without explicitly lifting degeneracy of the classical ground-state manifold
we are able to provide deeper insight into the quantum ObD effect.
More specifically, we shed light on the mechanism by which the choice is made between
${\bf q}\!=\!0$ and $\sqrt{3}\times\!\sqrt{3}$ ordered patterns  in KGAFMs
and present a rare example of the situation when quantum ObD defies the general trend
and yields the ground state that is different from the one favored by  thermal fluctuations.

% ------------------------------------------------------------------------------
% ------------------------------------------------------------------------------

On the technical side, we
take advantage of the fact that the so-called ``flat mode,'' the branch of localized linear spin-wave excitations
which has zero energy in the Heisenberg limit, see Fig.~\ref{fig:states}(c),
becomes gaped for $\Delta\!<\!1$ with $\varepsilon_{\bf k}\!\propto\!\sqrt{1-\Delta}$,  see Fig.~\ref{fig:states}(d).
Because of that, a controlled $1/S$ expansion becomes possible in the $XXZ$ KGAFM, allowing for a detailed
investigation of the quantum selection of the ordered state \cite{footnote_Dz}.

Another method that allows for an effective treatment of the highly-degenerate frustrated spin systems is the
real-space perturbation theory (RSPT). Applied to the KGAFMs, it operates directly within
the manifold of classical 120$^\circ$  states and, by analyzing terms of various order of the perturbation,
creates an intuitively transparent real-space hierarchy of effective couplings that are responsible
for the ground-state selection.
As we show below, it is the convolution of the two methods, $1/S$ expansion and RSPT, which is especially insightful.

% ------------------------------------------------------------------------------
\begin{figure}[t]
\includegraphics[width=0.99 \columnwidth]{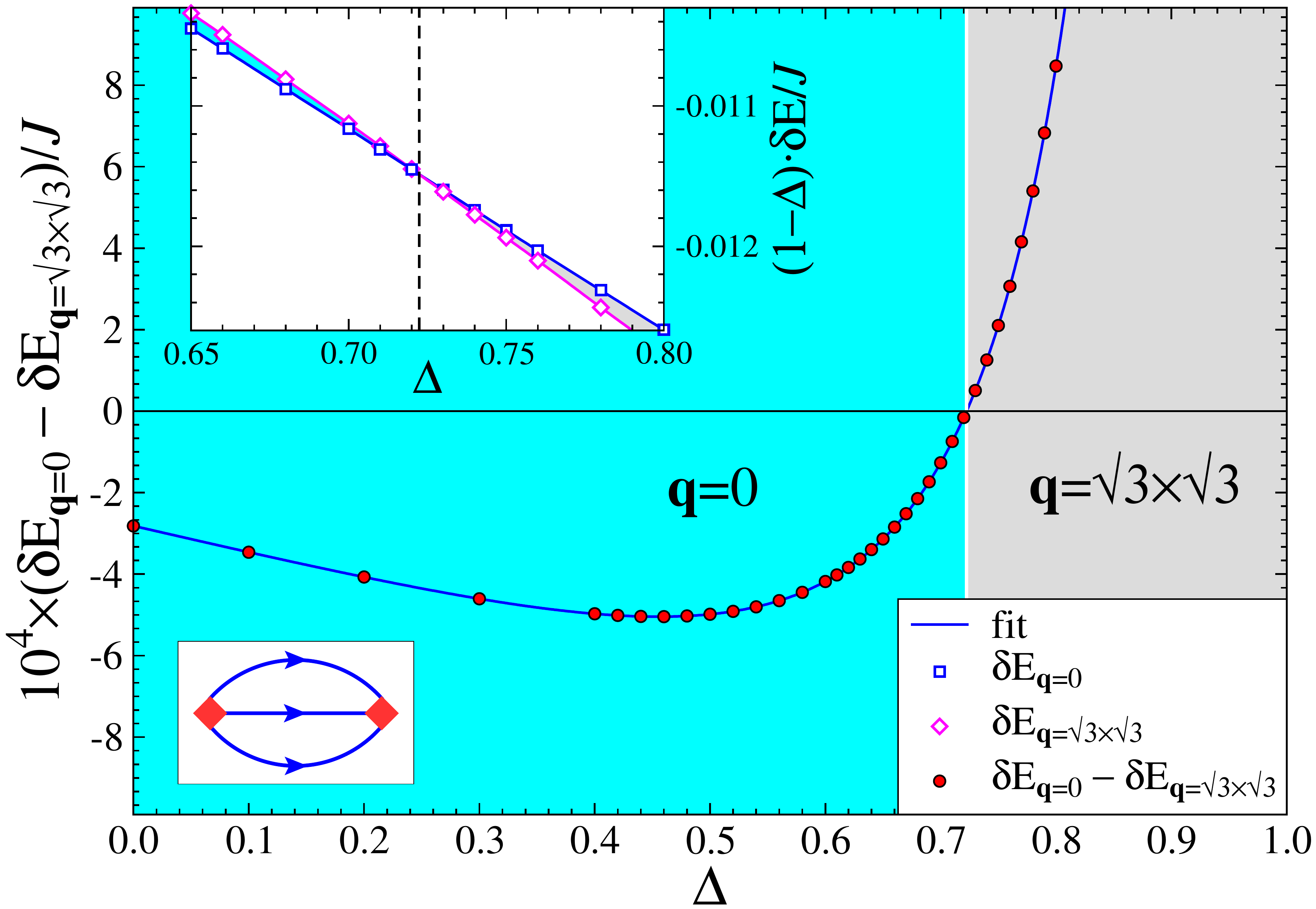}
\caption{(Color online) Difference of the ground-state energies (\ref{E})
of the ${\bf q}\!=\!0$ and $\sqrt{3}\times\!\sqrt{3}$ states, per spin. Upper inset:  energy correction $\delta E^{(3)}$
for the ${\bf q}\!=\!0$ (squares) and $\sqrt{3}\times\!\sqrt{3}$ (diamonds) states. Dashed line marks
the transition. Lower inset: diagram for  $\delta E^{(3)}$ term in the energy expansion.}
\label{fig:energy}
\end{figure}
% ------------------------------------------------------------------------------

% ==============================================================================
{\it Non-linear spin-wave theory (SWT).}---%
% ==============================================================================
For any  ordered state  from the coplanar 120$^\circ$ manifold one can rewrite (\ref{Ha})
in a rotating local basis as
\begin{eqnarray}
\hat{\cal H} &=& J \sum_{\langle ij\rangle} \Bigl( \Delta S_i^yS_j^y +
\cos \theta_{ij} \left(S_i^xS_j^x + S_i^z S_j^z\right) \nonumber\\
\label{H}
&&\phantom{J \sum_{\langle ij\rangle} \Bigl(}+\sin \theta_{ij} \left(S_i^zS_j^x - S_i^x S_j^z\right) \Bigr)\ ,
\end{eqnarray}
\vskip -0.4cm\noindent
where $\theta_{ij}\!= \!\theta_i \!-\! \theta_j$. % is the angle between two neighboring spins.
%This form makes it explicit
Clearly, it is only the last term in (\ref{H})  which is able to distinguish between different 120$^\circ$ spin configurations
by virtue of containing  $\sin\theta_{ij}\!= \!\pm\sqrt{3}/2$ for the clockwise
or counterclockwise spin rotation.
This term corresponds to the non-linear, cubic coupling of spin-waves and does not contribute to the linear SWT.
Consider the $1/S$ expansion of the ground-state energy
\begin{eqnarray}
E=E_{\rm cl}+\langle {\cal H}_2\rangle+\langle {\cal H}_4\rangle+\delta E^{(3)}+\dots,
\label{E}
\end{eqnarray}
where the first term is the classical energy ${\cal O}(S^2)$, the second is the
linear SWT correction ${\cal O}(S)$, and the last two are the contribution of the quartic and cubic terms,
both ${\cal O}(1)$. It is easy to see from (\ref{H}) that  quartic terms are also unable to differentiate between
120$^\circ$ structures, leaving the cubic term as a sole source of the quantum ObD effect to this $1/S$ order.
The energy correction from the cubic terms
is represented by the diagram in the lower inset of Fig.~\ref{fig:energy} and is given by
\begin{eqnarray}
\delta E^{(3)}=-\frac{1}{6N}\sum_{\nu\mu\eta}\sum_{{\bf q},{\bf k}}
\frac{|V^{\nu\mu\eta}_{{\bf q},{\bf k},-{\bf k}-{\bf q}}|^2}
{\varepsilon^\nu_{\bf q} + \varepsilon^\mu_{\bf k} + \varepsilon^\eta_{-{\bf k}-{\bf q}}}\, ,
\label{dE3}
\end{eqnarray}
where $\nu,\mu,\eta$ numerate spin-wave branches with harmonic energies
$\varepsilon^\alpha_{\bf k}$ and the cubic vertex comes from the anharmonic part of the spin-wave Hamiltonian
\begin{eqnarray}
\hat{\cal H}_3=\frac{1}{3!}\sum_{\nu\mu\eta}\sum_{{\bf q},{\bf k}}
V^{\nu\mu\eta}_{{\bf q},{\bf k},-{\bf k}-{\bf q}}\, b^\dag_{\nu,{\bf q}}b^\dag_{\mu,{\bf k}} b^\dag_{\eta,-{\bf k}-{\bf q}}+
\mbox{H.c.}
\label{dH3}
\end{eqnarray}
As is clear from  previous discussion, this vertex has \emph{different} form for different ordered structures
and should be obtained from the spin-wave expansion for each specific 120$^\circ$ spin pattern.

For the linear SWT of the $XXZ$ model (\ref{H}), we generalize the approach of \cite{Harris92}, which has
suggested a two-step diagonalization procedure consisting of the unitary transformation of the unit-cell bosons
followed by the Bogolyubov transformation for each mode.
With that we are able to obtain cubic vertices (\ref{dH3}) for the ${\bf q}\!=\!0$ and $\sqrt{3}\times\!\sqrt{3}$ states
in a fully analytic and elegant form \cite{suppl}, which permit high-accuracy numerical integration 
in (\ref{dE3}) and allow to study quantum ObD effect.
The results of such calculations are presented in Fig.~\ref{fig:energy}.

Our main result is the quantum selection of the ${\bf q}\!=\!0$ state over the $\sqrt{3}\times\!\sqrt{3}$
counterpart for anisotropy values extending from the $XY$ limit,
$\Delta\!=\!0$, to the transition point $\Delta_c\!\approx\!0.72235$.
This is contrary to the
common belief that quantum fluctuations follow the same selection trend as thermal ones.\cite{footnote_1}
Indeed, the asymptotic selection of the $\sqrt{3}\times\!\sqrt{3}$ magnetic structure
by thermal fluctuations for the classical KGAFM in both the Heisenberg 
\cite{Harris92,Reimers93,Henley09} and the $XY$  limits \cite{Huse92,Korshunov02,Rzchowski97} shows no
change in the ordering pattern as a function of $\Delta$ in contrast to the behavior of the
quantum model in Fig.~\ref{fig:energy}.
Although the $1/S$ energy correction diverges
as $(1-\Delta)^{-1}$,  signifying a failure of the expansion for $\Delta\!\rightarrow\!1$,
our results leave little doubt that the $\sqrt{3}\times\!\sqrt{3}$ state should remain
the ground state in the entire range $\Delta_c\!<\!\Delta\!\leq\!1$. Previously, a self-consistent
spin-wave treatment of the Heisenberg limit \cite{Chubukov92} has provided an indirect evidence in
favor of the $\sqrt{3}\times\!\sqrt{3}$ ground state for $S\!\gg\! 1$. Here this result
is strongly implied by a direct calculation of the ground-state energy. Lastly, we observe that
the energy gain from the quantum ObD effect  is only a fraction of $10^{-3}J$ per spin.

% ==============================================================================
{\it Real-space perturbation theory.}---%
% ==============================================================================
What is the mechanism of quantum selection of the ground state? We address this question
using the RSPT \cite{Long89,Heinila93,Canals04,Bergman07}. This approach 
divides the Hamiltonian  (\ref{H})  into an unperturbed part
$\hat{\cal H}_0 \!=\! h \sum_i \left(S-S_i^z\right)$, describing  spin fluctuations
in the local field  $h\!=\!2JS$, and perturbation $\hat{V}$, which couples 
 fluctuations on adjacent sites. Then, the standard perturbation theory is used 
to calculate  quantum corrections to the classical ground-state energy. The coupling 
between spin fluctuations contains four terms
$\hat{V} \!=\!  \sum_{i,j} (\hat{V}^{ij}_{1} \!+\! \hat{V}^{ij}_{2}
\!+\! \hat{V}^{ij}_{3} \!+\! \hat{V}^{ij}_{4} )$
\begin{eqnarray}
&&\hat{V}^{ij}_{1} =  -A_+
\bigl(S_i^+S_j^+ + \mbox{H.c.} \bigr), \quad
\hat{V}^{ij}_{2}  =  2A_-
S_i^+S_j^-,\quad \
\label{H_RSPT}
\\
&& \hat{V}^{ij}_{3} = -B_{ij}\,
\delta S_i^z\,\bigl(S_j^++S_j^-\bigr),\quad
\hat{V}^{ij}_{4}  =  -C \, \delta S_i^z\, \delta S_j^z\,, \nonumber
\end{eqnarray}
where we introduce $\delta S_i^z\!=\!S\!-\!S_i^z$,  $A_\pm \!=\!J\left(\Delta\!\pm\!1/2\right)/8$,
$B_{ij}\!=\! J\sin\theta_{ij}/2$, $C\!=\!J/4$,  and keep 
$\sin\theta_{ij}\!=\!\pm\sqrt{3}/2$ in $\hat{V}_3$ explicit, see \cite{suppl} for details.
The first three terms in (\ref{H_RSPT}) can be referred to as  double spin-flip,
spin-flip hopping, and  single spin-flip, the latter being a descendant of the cubic term  (\ref{H}).
As in the $1/S$ expansion, this  is the only term which is sensitive to the 120$^\circ$ pattern
and, therefore, is the key to the  selection of the ground state.

% ------------------------------------------------------------------------------
\begin{figure}[t]
\includegraphics[width=0.99 \columnwidth]{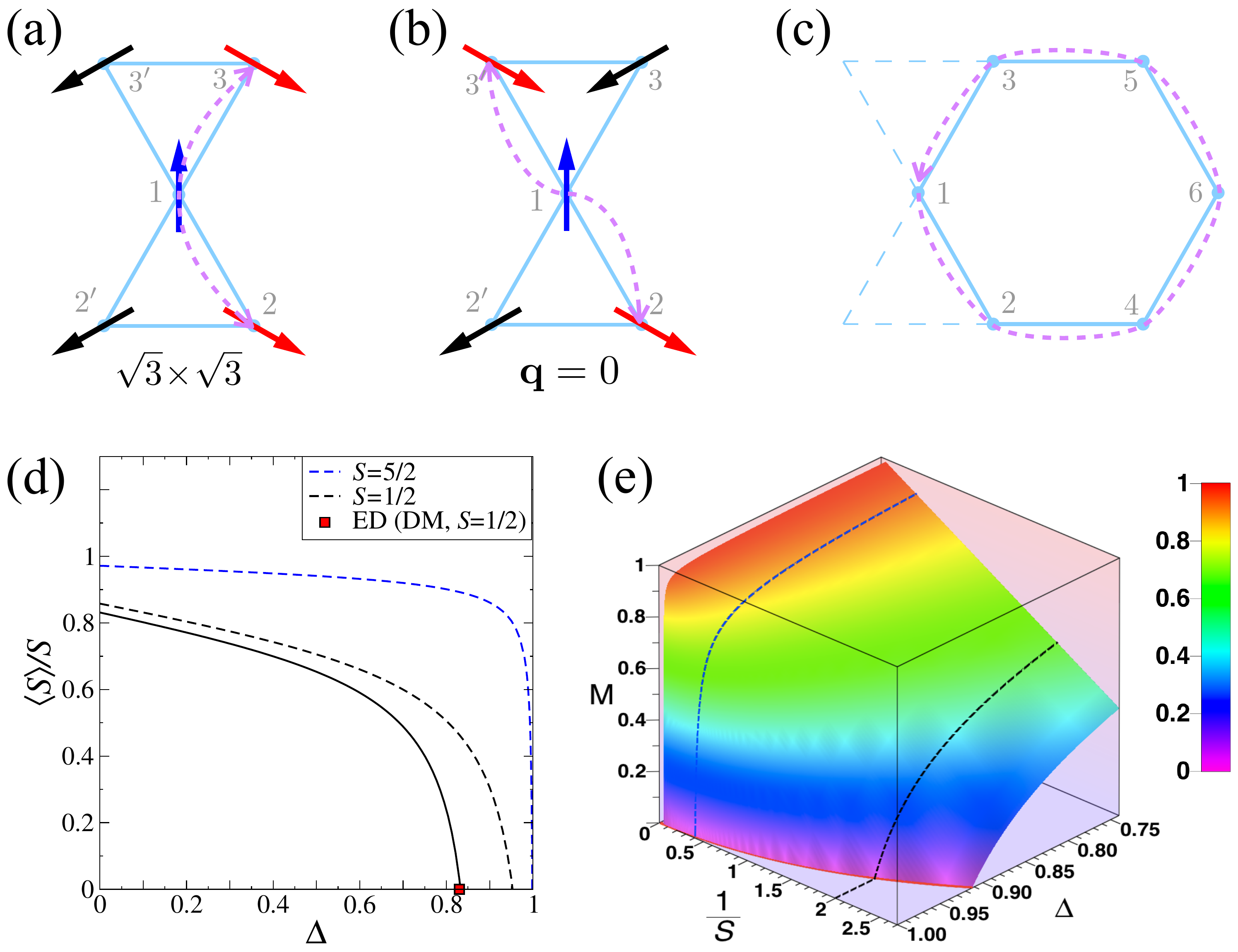}
\caption{(Color online) (a) and (b): schematics of the symmetry related  processes of the 4th order.
(c) Topologically non-trivial  path of the 7th order. (d) Magnetization
$M\!=\!\langle S\rangle/S$ vs $\Delta$ in linear SWT for $S\!=\!1/2$ and $S\!=\!5/2$, dashed lines.
Solid line, a sketch of $M(\Delta)$ for the case of DM interaction. (e) Linear SWT result for $M(S,\Delta)$. }
\label{fig:RSPT}
\end{figure}
% ------------------------------------------------------------------------------

Since every term in the energy expansion corresponds to
a finite cluster of spins coupled by perturbations in (\ref{H_RSPT})
and since the classical ground state is a vacuum for spin flips,
contributions that are relevant to lifting the ground-state degeneracy must begin and end
with a double spin-flip and must also contain a pair of single spin-flips.
The first process of such kind appears in the fourth order,
an example given by the operator sequence
$\hat{V}_{1}^{12} \!\to \! \hat{V}_{3}^{13} \!\to\! \hat{V}_{3}^{12}\!\to \! \hat{V}_{1}^{13}$
shown in  Fig.~\ref{fig:RSPT}(a).
The respective energy shift depends on the mutual orientation of ${\bf S}_2$
and ${\bf S}_3$ because $\delta E^{(4)} \!\propto \!\sin \theta_{12} \sin\theta_{13}$.
However, an obvious symmetry leaves the degeneracy intact at this order of expansion,
because for any coupling between ${\bf S}_2$ and ${\bf S}_3$ there is a
``mirror'' process that couples identically ${\bf S}_2$ with ${\bf S}_{3^\prime}$,
see Fig.~\ref{fig:RSPT}(b), providing the same energy gain to both the
$\sqrt{3}\times\!\sqrt{3}$ and the ${\bf q}\!=\!0$ states.

Generalizing this trend, we conclude that  the degeneracy-lifting terms 
must correspond to linked clusters of a non-trivial topology, with the smallest cluster consisting of a hexagon loop
 and generated by the seventh-order process depicted in Fig.~\ref{fig:RSPT}(c). One example of
the operator sequence is given by 
$\hat{V}_{1}^{24}\!\to\!\hat{V}_{3}^{21}\!\to\!\hat{V}_{3}^{13}\!\to\!
\hat{V}_{1}^{12} \!\to\! \hat{V}_{1}^{56} \!\to\! \hat{V}_{1}^{46}\!\to\!\hat{V}_{1}^{35}$ 
and contains five double-flips and two single-flips.
This type of processes yields the only relevant seventh-order contribution at $\Delta\!=\!1/2$,
for which the amplitude $A_-$ of the spin-flip hopping $\hat{V}_2$ (\ref{H_RSPT}) vanishes 
together with the rest of the degeneracy-lifting terms.
The energy correction at $\Delta\!=\!1/2$  corresponds
to an effective antiferromagnetic coupling between the second-neighbor spins ${\bf S}_2$ and ${\bf S}_3$, 
$\delta E^{(7)} \sim + \sin\theta_{12}\sin\theta_{13}$ \cite{suppl}, favoring the ${\bf q}\!=\!0$ state. 

Moreover, one can show that for $\Delta\!<\!1/2$ all relevant seventh-order
processes have the same sign  and also favor  the ${\bf q}\!=\!0$ state. For $\Delta\!>\!1/2$, some of the 
 terms switch sign. This implies that
the transition to the $\sqrt{3}\times\!\sqrt{3}$ state can
 take place only at some $\Delta_c\!>\!1/2$, in agreement with the the non-linear SWT result $\Delta_c\!\approx\!0.72$.

There are close parallels between the non-linear SWT and the real-space approach.
Although the degeneracy-lifting contribution in the RSPT is of the seventh order, it is still
of the second order in cubic terms, same as in the non-linear SWT (\ref{dE3}).
More importantly, the high order of the relevant perturbation processes explains the 
smallness of the quantum ObD effect. Essentially, the RSPT is an expansion in
$1/z$, where $z$ is the coordination number, which gives the right order-of-magnitude  
estimate for the seventh-order effect
$\delta E\!\sim\! 10^{-4}J$. 
A  more careful calculation  using the actual perturbation terms in (\ref{H_RSPT}) and    
combinatorial factors of  different  processes of  seventh order  gives a similar answer \cite{suppl}.
Our conclusion on the topological nature of the effective exchange responsible for the ground-state selection
also makes it extremely unlikely that a state with an extended unit cell can compete
with the ones considered in this work.

% ==============================================================================
{\it Phase diagram.}---%
% ==============================================================================
We now construct the phase diagram of the $XXZ$ KGAFM (\ref{Ha}) as a function of
anisotropy $\Delta$ and spin $S$. For that, we calculate the ordered moment within the harmonic SWT
approximation, $\langle S\rangle\!=\!S\!-\!\langle a_i^\dag a_i\rangle$, to map out the extent of
the magnetically ordered state.
Because of the degeneracy of classical 120$^\circ$ states, harmonic spin-wave spectrum is identical in
all of them and yields the same result.
%\cyan{This is just another way of saying that the ground-state selection among the ordered states requires
%non-linear terms.} 
 Here we simply estimate  stability of the Ne\'{e}l order
with respect to the ``diagonal'' quantum fluctuation for a given state.
While this analysis completely neglects  the ``off-diagonal''  tunneling  within the
manifold, such processes should be exponentially suppressed for larger spins \cite{Henley93}.

\begin{figure}[t]
\includegraphics[width=0.99 \columnwidth]{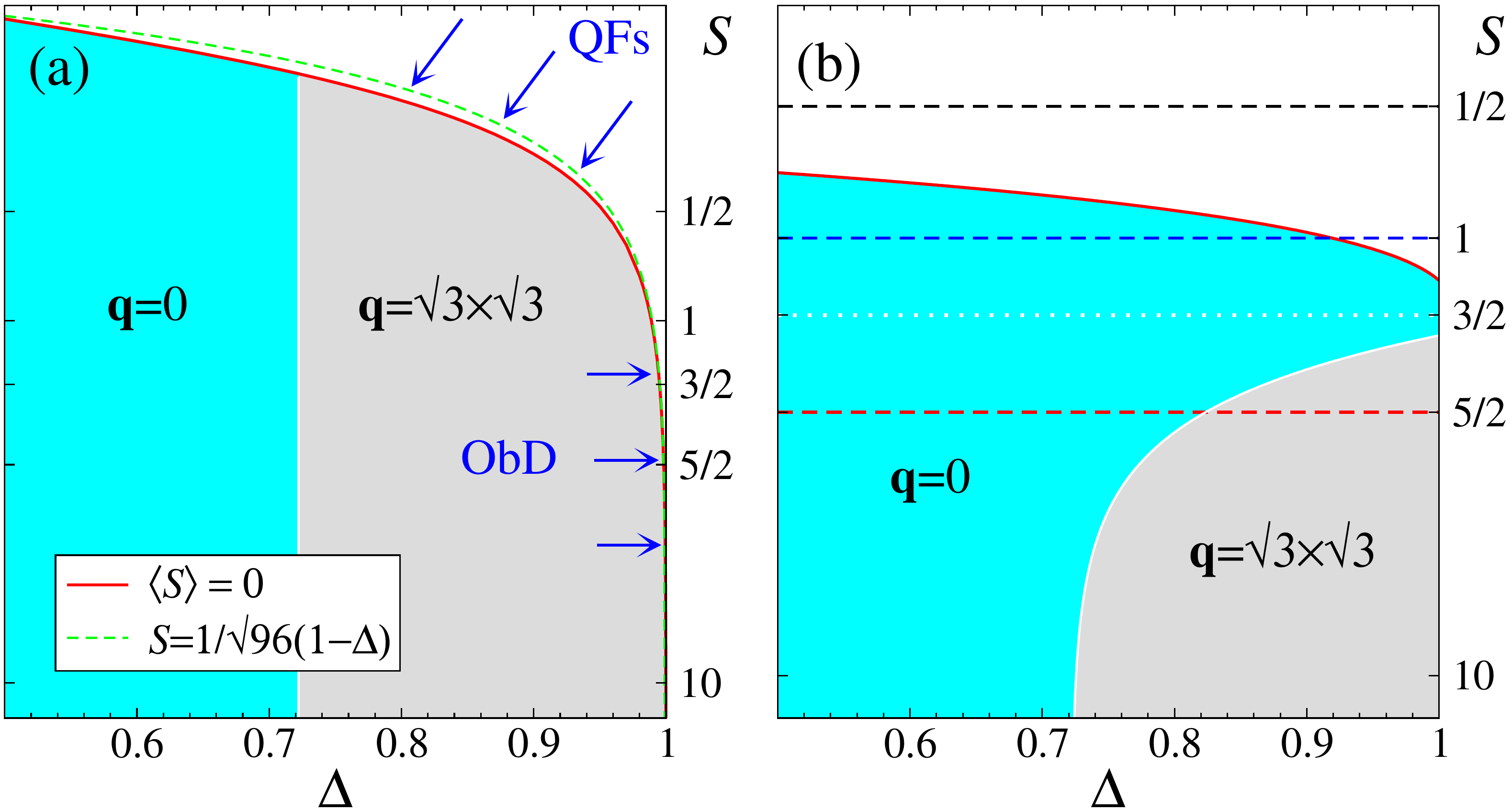}
\caption{(Color online) (a) $S\!-\!\Delta$ phase diagram, $S$ is on the logarithmic scale. Solid line is
the Ne\'{e}l order boundary $\langle S\rangle\!=\!0$ from the linear SWT, dashed line is its asymptotic approximation.
Vertical boundary between ${\bf q}\!=\!0$ and $\sqrt{3}\times\!\sqrt{3}$ states is from the non-linear SWT.
%Inset: same on the log-log scale.
(b) Tentative $S\!-\!\Delta$ phase diagram. Suppression of magnetic
order by quantum fluctuations (QFs) and quantum ObD near the Heisenberg limit
are suggested. Horizontal lines are cuts for different  values of the spin.}
\label{fig:phd}
\end{figure}

Figure \ref{fig:RSPT}(d) shows magnetization $M\!=\!\langle S\rangle/S$ vs
$\Delta$ for two representative values of the spin.  Ne\'{e}l state is stabilized  already
at rather small $1\!-\!\Delta_c^\prime\!\approx\! 0.05$ for $S\!=\!1/2$
and $1\!-\!\Delta_c^\prime\!\approx\! 0.002$ for $S\!=\!5/2$. Considering spin $S$ as a continuous
variable, we plot $M(S,\Delta)$ in Fig.~\ref{fig:RSPT}(e) where  dashed lines are
the same as in Fig.~\ref{fig:RSPT}(d) and the color is for the magnitude of $M$. The  $M\!=\!0$
curve is the Ne\'{e}l order phase boundary in the $S\!-\!\Delta$ plane, see also Fig.~\ref{fig:phd}(a).
A simple algebra yields an asymptotic expression for it, $1\!-\!\Delta_c^\prime\!\approx\!(96S^2)^{-1}$,
which agrees exceedingly well with the results of numerical integration \cite{suppl}.

In Fig.~\ref{fig:RSPT}(d) we also sketch results of the Exact Diagonalization (ED)  for
$S\!=\!1/2$ KGAFM with the out-of-plane DM term \cite{Cepas08},
which selects  ${\bf q}\!=\!0$ ground state  but yields
harmonic Hamiltonian identical to the $XXZ$ case with rescaling
$1\!-\!\Delta\Leftrightarrow\sqrt{3}D_z$ \cite{Yildirim06,Canals03,footnote_Dz}.
Since the DM term suppresses tunneling processes within the manifold,  it is reasonable
to compare ED with SWT results to evaluate the accuracy of the SWT  Ne\'{e}l order boundary.
For the latter, one can see a qualitative agreement with ED and a quantitative exaggeration of the
extent of the ordered phase, expected for the SWT approach.

We now combine our SWT results in Fig.~\ref{fig:phd}(a), which shows the
$S\!-\!\Delta$ phase diagram.
The solid line is the linear SWT result for the Ne\'{e}l order boundary
$\langle S\rangle\!=\!0$, see also Fig.~\ref{fig:RSPT}(e), and the dashed line is its asymptotic
approximation mentioned above.
As we discussed, the harmonic treatment  gives a good qualitative idea for the phase boundary between
magnetically ordered and disordered phases, but \emph{does not} specify which
of the 120$^\circ$  Ne\'{e}l states is chosen.
We infer this information from the non-linear SWT results in  Fig.~\ref{fig:energy} and complete
our perturbative $S\!-\!\Delta$ phase diagram by adding the boundary between
${\bf q}\!=\!0$ and $\sqrt{3}\times\!\sqrt{3}$ states. 
% This boundary is vertical  because the corresponding term in the $1/S$ expansion of energy (\ref{dE3}) is independent of $S$.

% ------------------------------------------------------------------------------
% ------------------------------------------------------------------------------

There are two trends that are not included in this phase diagram and are beyond
the methods used in the current work.
The non-linear SWT approach is a perturbative treatment of the quantum ObD effect, which fails in the
vicinity of the Heisenberg limit.
However,  it is known that quantum ObD should extend  Ne\'{e}l-ordered region
of the phase diagram to the $\Delta\!=\!1$ axis for larger spin values. It has been argued by
the self-consistent version of SWT \cite{Chubukov92} that the $\sqrt{3}\times\!\sqrt{3}$ is the ground state
of the Heisenberg KGAFM for $S\!\gg\!1$.

The other trend is the suppression of the Ne\'{e}l  order by quantum fluctuations for smaller spins, leading to
the growth of the non-magnetic region of the phase diagram.  As was argued recently by several groups using numerical
approaches \cite{Lauchli_unpub,Steve_unpub,chinese_arxiv}, the $S\!=\!1/2$ $XXZ$ KGAFM remains in a spin-liquid state
for the entire range of $\Delta\!\leq\!1$. Our results for $S\!=\!1/2$ case in Fig.~\ref{fig:phd}(a) are, therefore,
inadequate, most likely because of the neglect of the tunneling between different states in 120$^\circ$ manifold.
%\cyan{However, for larger spins the tunneling processes are suppressed exponentially  and our results can be expected
%to hold.}

In order to capture some of these trends we modify the mean-field condition $\langle S\rangle\!=\!0$ used above
by including the self-consistently renormalized spin-wave dispersion of the ``flat mode'' for the Heisenberg limit
from \cite{Chubukov92}. While this is not an entirely rigorous procedure, it should provide a reasonable
estimate on the extent of the region of stability due to quantum ObD for $\Delta\!=\!1$. The resulting values for the
``critical'' $S_c$, above which the system orders magnetically, come out as  $S_c^{{\bf q}=0}\!\approx\!0.17$
and $S_c^{\sqrt{3}\times\sqrt{3}}\!\approx\!0.18$. While, obviously, this is another case of
 quantitative exaggeration of the extent of the ordered phase by an SWT approach,
this estimate makes it extremely unlikely that the Heisenberg KGAFM with $S\!\agt\!1$ will be
magnetically disordered. In fact,  recent numerical work \cite{Gotze11} has indicated that the Heisenberg KGAFMs
with $S\!\geq\!3/2$ all order in a $\sqrt{3}\times\!\sqrt{3}$ configuration.

Combining these trends, we propose a tentative $S\!-\!\Delta$ phase diagram of the nearest-neighbor
$XXZ$ KGAFM model  in Fig.~\ref{fig:phd}(b).
In the Heisenberg limit, for  larger values of spin the ground state is
$\sqrt{3}\times\!\sqrt{3}$ state, which switches
to ${\bf q}\!=\!0$ upon reducing $\Delta$. For $S\!=\!1$ the same trajectory begins with the
magnetically disordered state and the system enters directly into the ${\bf q}\!=\!0$ state. As  shown by the
recent numerical results, $S\!=\!1/2$ remains quantum disordered for the entire range of $\Delta$.
Finally, there may, or may not, exist an intermediate value of spin for which
 Heisenberg limit is already in the ${\bf q}\!=\!0$ domain and no transition occurs
versus $\Delta$. While predictions of this work are firm for the
larger values of  spin, the ultimate answer on the exact sequence of phases for smaller spins should
be sought via numerical approaches.

% ==============================================================================
{\it Conclusions.}---%
% ==============================================================================
By advancing the non-linear  $1/S$ expansion  and  the real-space perturbation theory
we investigated quantum order-by-disorder selection of the ground state of
the  nearest-neighbor $XXZ$ antiferromagnet on the kagom\'{e} lattice.
We demonstrated that the order selection
is generated by topologically non-trivial  tunneling processes,
presented a strong evidence of the rare case of quantum and thermal fluctuations favoring  different ground states,
proposed a tentative $S\!-\!\Delta$ phase diagram of the model, and suggested
further  studies.

We acknowledge useful discussions with  C. Batista, F. Becca, A. V. Chubukov, G. Jackeli,
A. L\"auchli, R. Moessner, N. Perkins, S. Parameswaran, H. Tsunetsugu, and S. R. White.
Work by A.~L.~C. was supported by the U.S. Department of Energy, Office of Science, Basic Energy Sciences
under Award \# DE-FG02-04ER46174.  A.~L.~C. would like to thank Aspen Center for Physics, where part of this work
was done, for hospitality. The work at Aspen was supported in part by NSF Grant No. PHYS-1066293.

% ==============================================================================

\newpage
% ==============================================================================
%---------------------------------------------------------------------------
%\begin{widetext}
\onecolumngrid
%---------------------------------------------------------------------------
\begin{center}
%{\large\bf Quantum Selection of Order in $XXZ$ Kagom\'{e} Antiferromagnet: \\ Supplemental Material}\\ 
{\large\bf Quantum Selection of Order in an $XXZ$ Antiferromagnet on a Kagom\'{e} Lattice: \\ Supplemental Material}\\ 
\vskip0.35cm
A. L. Chernyshev$^1$ and M. E. Zhitomirsky$^2$\\
\vskip0.15cm
{\it \small $^1$Department of Physics and Astronomy, University of California, Irvine, California
92697, USA\\
$^2$Service de Physique Statistique, Magn\'etisme et Supraconductivit\'e,\\
UMR-E9001 CEA-INAC/UJF, 17 rue des Martyrs, 38054 Grenoble Cedex 9, France}\\
{\small (Dated: October 24, 2014)}\\
\vskip 0.1cm \
\end{center}
%---------------------------------------------------------------------------
\twocolumngrid
%\end{widetext}
%---------------------------------------------------------------------------
% ==============================================================================

% ==============================================================================
\section{Spin-wave theory}
% ==============================================================================
\subsection{Spin Hamiltonian}

We consider a kagom\'e-lattice antiferromagnet with anisotropic $XXZ$ exchange interactions
\begin{equation}
\hat{\cal H} = J \sum_{\langle ij\rangle} \Bigl( S_i^xS_j^x + S_i^yS_j^y
+ \Delta S_i^z S_j^z \Bigr) \ ,
\label{Has}
\end{equation}
where summation is over bonds, $i,j$ numerate the sites of the kagom\'e lattice, and
anisotropy is assumed to be of the  easy-plane type, $0\leq\Delta \leq 1$. In
a semiclassically ordered state spins of the kagom\'e-lattice antiferromagnet form
a coplanar 120$^\circ$ structure in the $xy$ plane. Transforming to a rotating local
basis we can rewrite \eqref{Has} as
\begin{eqnarray}
\hat{\cal H} &=& J \sum_{\langle ij\rangle} \Bigl( \Delta S_i^yS_j^y +
\cos \theta_{ij} \left(S_i^xS_j^x + S_i^z S_j^z\right) \nonumber\\
\label{Hs}
&&+\sin \theta_{ij} \left(S_i^zS_j^x - S_i^x S_j^z\right) \Bigr)
= J \sum_{\langle ij\rangle} {\bf S}_i \otimes{\bf S}_j
\ ,
\end{eqnarray}
where $\theta_{ij}= \theta_i - \theta_j$ is an angle between two neighboring spins and we have
introduced ``matrix'' product of spins $\otimes$ as a shorthand notation.
% -----------------------------------------------------------------------------
\begin{figure}[b]
\includegraphics[width=0.5\columnwidth]{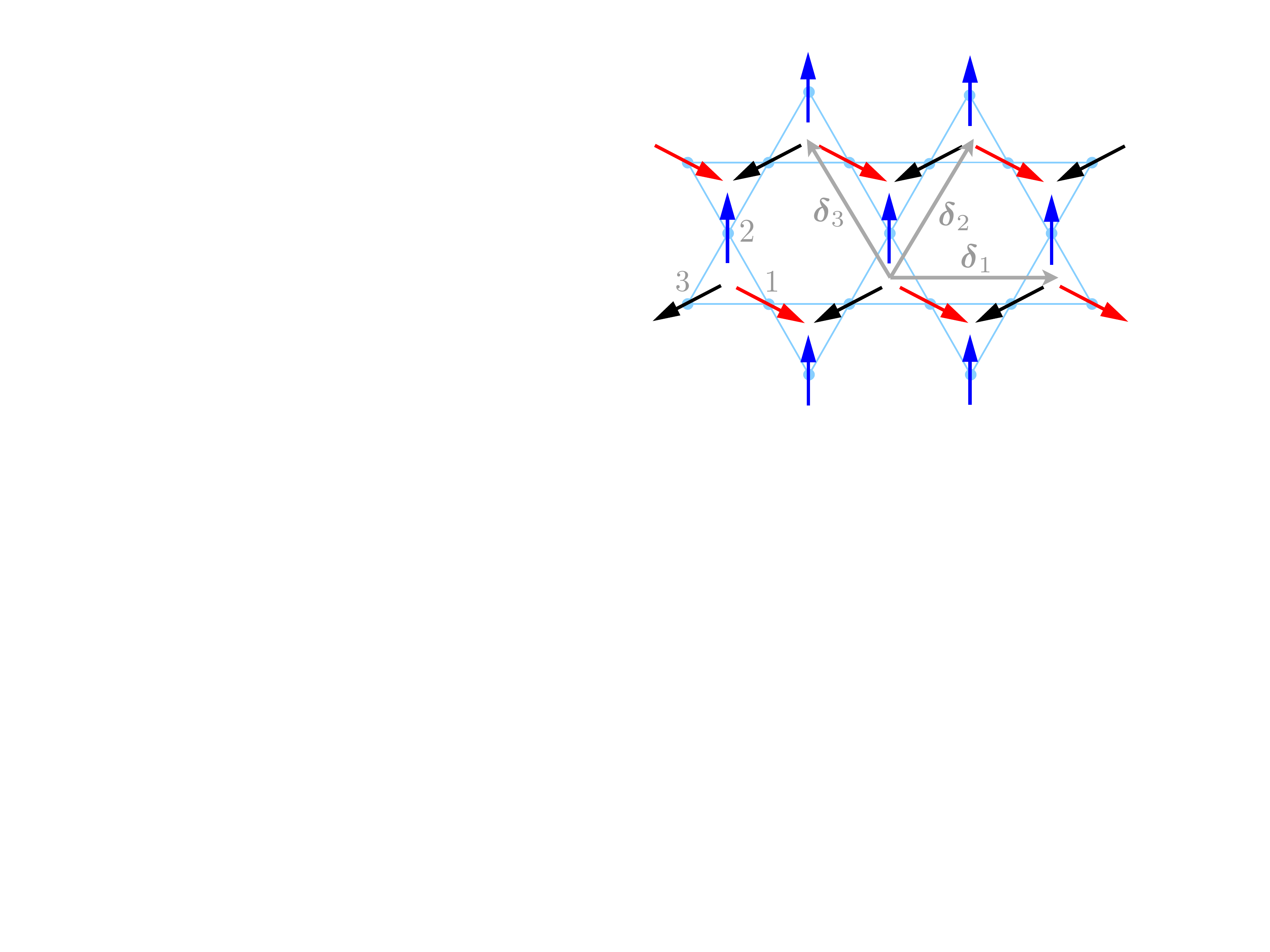}
\caption{
Numbering of sites within the unit cell and primitive vectors of the kagom\'e lattice.
}
\label{basis}
\end{figure}
% ------------------------------------------------------------------------------

We choose the unit cell of the kagom\'e lattice as an up-triangle with three atoms in the positions
\begin{equation}
\bm{\rho}_1 = (0,0)\, , \
\bm{\rho}_2 = \biggl(-\frac{1}{4},\frac{\sqrt{3}}{4}\biggr)\, , \
\bm{\rho}_3 = \biggl(-\frac{1}{2},0\biggr)\,.
\end{equation}
All distances are given in units of $2a$ where $a$ is the interatomic distance.
The corresponding Bravais lattice is a triangular lattice with the primitive vectors
\begin{eqnarray}
\bm{\delta}_1 = (1,0)\, , \
\bm{\delta}_2 = \biggl(\frac{1}{2}, \frac{\sqrt{3}}{2}\biggr)\, , \
\bm{\delta}_3 = \bm{\delta}_2-\bm{\delta}_1\, ,
\end{eqnarray}
such that $\bm{\rho}_2 = \frac{1}{2}\bm{\delta}_3$ and
$\bm{\rho}_3 = -\frac{1}{2}\bm{\delta}_1$, see Fig.~\ref{basis}.
Then, changing the lattice sum to the sum over the unit cells and atomic index,
$i\rightarrow\{\alpha,\ell\}$, with $\alpha=1,2,3$ numerating atoms
within the unit cell, Hamiltonian \eqref{Hs} becomes
\begin{eqnarray}
\label{Hls}
\hat{\cal H} &=& J \sum_\ell {\bf S}_{1,\ell}\otimes\left({\bf S}_{2,\ell} + {\bf S}_{2,\ell-3}\right)  \\
&&+ {\bf S}_{1,\ell}\otimes\left({\bf S}_{3,\ell} + {\bf S}_{3,\ell+1}\right) +
{\bf S}_{2,\ell}\otimes\left({\bf S}_{3,\ell} + {\bf S}_{3,\ell+2}\right) \, ,\nonumber
\end{eqnarray}
where the product ${\bf S}_{\alpha,\ell}\otimes {\bf S}_{\alpha',\ell'}$ is according to (\ref{Hs}),
and $\ell\pm n\equiv{\bf R}_\ell \pm\bm{\delta}_n$ with the coordinate of the unit cell
${\bf R}_\ell= m_1\bm{\delta}_1 + m_2\bm{\delta}_2 + m_3\bm{\delta}_3$.

% ==============================================================================
\subsection{Linear spin-wave theory}
% ==============================================================================

Following the approach of Ref.~\cite{Harris92s},  we introduce  Holstein-Primakoff
representation for spin operators in the local basis in \eqref{Hs} and
\eqref{Hls} and, keeping only quadratic  terms, obtain harmonic Hamiltonian
for the three species of bosons, $a_{\alpha,\ell}$($a^\dag_{\alpha,\ell}$).
With the subsequent Fourier transformation of bosonic operators performed according to
\begin{equation}
a_{\alpha,\ell} = \frac{1}{\sqrt{N}} \sum_{\bf k}  a_{\alpha,\bf k} \, e^{i{\bf k}{\bf r}_{\alpha,\ell}}
\end{equation}
where ${\bf r}_{\alpha,\ell}\! =\! \bm{\rho}_\alpha\! +\! {\bf R}_\ell$ and $N$ is the number of unit cells,
we obtain linear SWT Hamiltonian
\begin{eqnarray}
\hat{\cal H}_2 &=& 2 JS \sum_{{\bf k},\alpha\beta}\biggl\{\Bigl[\delta_{\alpha,\beta}+
\frac{\left(2\Delta-1\right)}{4}\,
\Lambda_{\alpha\beta}({\bf k})\Bigr] a_{\alpha,\bf k}^\dagger a_{\beta,\bf k} \nonumber\\
&&\phantom{2J}- \frac{(2\Delta+1)}{8}\,\Lambda_{\alpha\beta}({\bf k})\bigl(
a_{\alpha,\bf k}^\dagger a^\dagger_{\beta,-\bf k}+\textrm{h.c.}\bigr)\biggr\},\label{H2F}
\end{eqnarray}
where we introduce the matrix
\begin{equation}
\label{Mk}
\hat{\bm\Lambda}_{\bf k} =\left(\begin{array}{ccc}
0   &  c_3 & c_1 \\
c_3 &  0   & c_2 \\
c_1 &  c_2 &  0
\end{array}\right) \, ,
\end{equation}
and shorthand notations $c_n = \cos(q_n)$ with $q_n\!=\!{\bf k}\cdot\bm{\delta}_n/2$.

One can rewrite this Hamiltonian in the matrix form
\begin{equation}
\label{Hmatrix}
\hat{\cal H}_2 =  \sum_{\mathbf{k}>0}
\hat{X}^\dagger_{\bf k} \hat{\bf H}_{\bf k} \hat{X}_{\bf k} - 3JS\, ,
\end{equation}
with  the vector operator
\begin{equation}
\label{Xvector}
\hat{X}^\dagger_{\bf k} \!=\!
\bigl(a_{1,\bf k}^\dagger, a_{2,\bf k}^\dagger, a_{3,\bf k}^\dagger,
a_{1,-\bf k},a_{2,-\bf k},a_{3,-\bf k}\bigr)
\end{equation}
and the $6\!\times\! 6$ matrix $\hat{\bf H}_{\bf k}$
\begin{equation}
\hat{\bf H}_{\bf k} = 2J S\left(\begin{array}{cc}
\hat{\bf A}_{\bf k} &  \hat{\bf B}_{\bf k}  \\
\hat{\bf B}_{\bf k}  &    \hat{\bf A}_{\bf k}
\end{array}\right)\, ,
\end{equation}
where
\begin{equation}
\hat{\bf A}_{\bf k} =\hat{\bf I}+\frac{\left(2\Delta-1\right)}{4} \, \hat{\bm\Lambda}_{\bf k}, \ \
\hat{\bf B}_{\bf k} =-\frac{\left(2\Delta+1\right)}{4} \, \hat{\bm\Lambda}_{\bf k} \, ,
\label{AB}
\end{equation}
and $\hat{\bf I}$ being the identity matrix.

Because of an obvious commutativity of the matrices $\hat{\bf A}_{\bf k}$ and $\hat{\bf B}_{\bf k}$, their eigenvalues
are straightforwardly related to the eigenvalues of $\hat{\bf H}_{\bf k}$, and, in turn, are
determined by the eigenvalues of the matrix $\hat{\bm\Lambda}_{\bf k}$, so that
the spin-wave excitation energies are
\begin{equation}
\varepsilon_{\nu,\bf k} = 2JS\sqrt{A_{\nu,\bf k}^2-B_{\nu,\bf k}^2}
=  2JS \omega_{\nu,\bf k} \ ,
\label{Ek}
\end{equation}
with
\begin{equation}
A_{\nu,\bf k} =1+\frac{\left(2\Delta-1\right)}{4} \, \lambda_{\nu,\bf k}, \
B_{\nu,\bf k} =-\frac{\left(2\Delta+1\right)}{4} \, \lambda_{\nu,\bf k} \, .
\label{AB1}
\end{equation}
Thus, the problem of diagonalization of $\hat{\cal H}_2$ in (\ref{H2F}) is reduced to the
eigenvalue problem of $\hat{\bm\Lambda}_{\bf k}$ (\ref{Mk}).
From the characteristic equation for the matrix $\hat{\bm\Lambda}_{\bf k}$  one finds
\begin{equation}
|\hat{\bm\Lambda}_{\bf k} -\lambda | =
\left(\lambda +1\right) \left(\lambda^2 - \lambda - 2\gamma_{\bf k}\right) = 0 \ ,
\end{equation}
where $\gamma_{\bf k} \equiv c_1 c_2 c_3$ is introduced and factorization is performed with the help
of a useful identity
\begin{equation}
c_1^2 + c_2^2 + c_3^2 = 1 + 2 c_1 c_2 c_3\, .
%\cos^2 x + \cos^2y + \cos^2(x+y) = 1 + 2 \cos x \cos y \cos(x+y)
\nonumber
\end{equation}
Thus, the $\lambda$-eigenvalues are
\begin{equation}
\lambda_1 = -1 \ , \quad \lambda_{2(3),{\bf k}} = \frac{1}{2}\,\left(1 \pm \sqrt{1 + 8\gamma_{\bf k}}\right) \, ,
\label{lambda123}
\end{equation}
and one of the spin-wave excitations is completely dispersionless (``flat mode'')
\begin{equation}
\label{w1}
\varepsilon_{1,\bf k}  = 2JS \sqrt{3(1-\Delta)/2} \, ,
\end{equation}
whereas the other two are given by
\begin{equation}
\varepsilon_{2(3),\bf k}  = 2JS \sqrt{1-\Delta\gamma_{\bf k} -(1-\Delta)
\bigl(1 \pm \sqrt{1 + 8\gamma_{\bf k}}\,\bigr)/4}\, .\nonumber
\end{equation}
In the Heisenberg limit, $\Delta=1$, the flat mode has zero energy, while the other two
modes are degenerate
\begin{equation}
\varepsilon_{2(3),\bf k}  = 2JS \sqrt{1-\gamma_{\bf k}} \ .
\end{equation}

% ==============================================================================
\subsubsection{Two-step diagonalization}
% ==============================================================================

Following Harris {\it et al.} \cite{Harris92s}, the diagonalization of
$\hat{\bm\Lambda}_{\bf k}$ implies a two-step diagonalization procedure
of $\hat{\cal H}_2$ in (\ref{H2F}).
The eigenvectors ${\bf w}_{\nu}=\left(w_{\nu,1}({\bf k}),w_{\nu,2}({\bf k}),w_{\nu,3}({\bf k})\right)$
\begin{equation}
\hat{\bm\Lambda}_{\bf k} {\bf w}_{\nu} =\lambda_{\nu,\bf k} {\bf w}_{\nu}
\end{equation}
are given explicitly by
\begin{equation}
\label{wn}
{\bf w}_\nu ({\bf k})= \frac{1}{r_\nu} \,\left(\begin{array}{c}
c_1c_2 + \lambda_\nu c_3 \\
\lambda_\nu^2 - c_1^2 \\
c_1 c_3 + \lambda_\nu c_2
\end{array}\right)\, ,
\end{equation}
with $r_\nu = \sqrt{(c_1c_2 + \lambda_\nu c_3)^2 +
(\lambda_\nu^2 - c_1^2)^2 + (c_1 c_3 + \lambda_\nu c_2)^2}$.

These eigenvectors define a \emph{unitary} transformation from the Holstein-Primakoff
bosons to the new ones
\begin{equation}
d_{\nu,\bf k} = \sum_\alpha w_{\nu,\alpha}({\bf k})\, a_{\alpha,\bf k}\, , \ \ \
a_{\alpha,\bf k} = \sum_\nu w_{\nu,\alpha}({\bf k})\, d_{\nu,\bf k}\, ,
\label{linearT}
\end{equation}
such that  $\hat{\cal H}_2$ in (\ref{H2F}) is partially diagonalized
\begin{eqnarray}
\hat{\cal H}_2 &=& 2JS  \sum_{\nu,\bf k} \biggl( A_{\nu,\bf k} d_{\nu,\bf k}^\dagger d_{\nu,\bf k}\nonumber\\
&&\phantom{2JS  \sum_{\nu,\bf k}\biggl(}
- \frac{B_{\nu,\bf k}}{2}\left(d^\dagger_{\nu,\bf k} d^\dagger_{\nu,-\bf k} + \textrm{h.\,c.}\right)\biggr)
\,.
\label{H2P}
\end{eqnarray}
Finally, we apply the canonical Bogolyubov transformation for each individual species of $d$-boson
\begin{equation}
d_{\nu,\bf k} = u_{\nu\bf k}  b_{\nu,\bf k} + v_{\nu\bf k}  b^\dagger_{\nu,-\bf k} \,,
\label{BogolyubovT}
\end{equation}
with $u_{\nu\bf k}^2 -  v_{\nu\bf k}^2 = 1$ and 
\begin{equation}
v_{\nu\bf k}^2= \frac12\left(\frac{A_{\nu,\bf k}}{\omega_{\nu,\bf k}}-1\right),\quad
2u_{\nu\bf k}v_{\nu\bf k} = \frac{B_{\nu,\bf k}}{\omega_{\nu,\bf k}} \,,
\label{Bogolyubov_uv}
\end{equation}
to diagonalize (\ref{H2P}) completely with the eigenvalues (\ref{Ek}).
The importance of this two-step procedure will be apparent in the discussions of the non-linear terms.

% ==============================================================================
\subsection{Cubic terms}
% ==============================================================================

The non-linear $S_i^x S_j^z$ terms in  (\ref{Hs}) are the only ones
that are able to distinguish between different 120$^\circ$ spin configurations
by virtue of containing  $\sin\theta_{ij}\!= \!\pm\sqrt{3}/2$ for the clockwise
or counterclockwise spin rotation.
In the bosonic representation they yield cubic terms
\begin{equation}
\hat{\cal H}_3 = J\sqrt{\frac{S}{2}} \sum_{i,j} \sin\theta_{ij} \bigl( a_i^\dagger a_j^\dagger a_j +
\textrm{h.c.}\bigr)\, ,
\label{H3s}
\end{equation}
where $\theta_{ij}= \theta_i - \theta_j$ is an angle between two neighboring spins as before.
This results into anharmonic interaction of spin waves with the amplitudes which are
\emph{different}  for different ordered structures. Below we obtain cubic vertices for the
${\bf q}\!=\!0$ and $\sqrt{3}\times\!\sqrt{3}$ states.

For the ${\bf q}\!=\!0$  pattern (\ref{H3s}) can be rewritten as
\begin{equation}
\hat{\cal H}_3= -J\sqrt{\frac{3S}{2N}}\sum_{\alpha\beta,\bf k,q}
\epsilon^{\alpha\beta\gamma} \cos(q_{\beta\alpha})
a^\dagger_{\alpha,\bf q} a^\dagger_{\beta,\bf k} a_{\beta,\bf p}+ \textrm{h.\,c.},
\label{H30}
\end{equation}
where $\epsilon^{\alpha\beta\gamma}$ is the Levi-Civita
antisymmetric tensor, ${\bf p}={\bf k}+{\bf q}$,  and shorthand notations
$q_{\beta\alpha}={\bf q}\bm{\rho}_{\beta\alpha}$
 and $\bm{\rho}_{\beta\alpha} = \bm{\rho}_\beta - \bm{\rho}_\alpha$ are introduced.

Next we perform the two-step transformation. The unitary transformation (\ref{linearT}) yields
\begin{equation}
\hat{\cal H}_3 = -J\sqrt{\frac{3S}{2N}}\sum_{\bf k,q}\sum_{\nu\mu\eta}
F^{\nu\mu\eta}_{\bf qkp}\,
d_{\nu,\bf q}^\dagger d_{\mu,\bf k}^\dagger  d_{\eta,\bf p}  + \textrm{h.c.},
\label{H31}
\end{equation}
with the amplitude
\begin{equation}
F^{\nu\mu\eta}_{\bf qkp}=  \sum_{\alpha\beta}
\epsilon^{\alpha\beta\gamma}\cos(q_{\beta\alpha}) \,
w_{\nu,\alpha}({\bf q}) w_{\mu,\beta}({\bf k}) w_{\eta,\beta}({\bf p})\,.
\label{F0}
\end{equation}
The subsequent Bogolyubov transformation (\ref{BogolyubovT}) generates the ``source'', $b^\dag b^\dag b^\dag$,
and the ``decay'', $b^\dag b^\dag b$, terms.  For the $1/S$ expansion of the  groundstate energy
we need only the source terms
\begin{equation}
\hat{\cal H}_3  = \frac{1}{3!} \frac{1}{\sqrt{N}} \sum_{\bf k+p+q=0}
V^{\nu\mu\eta}_{\bf qkp}\,
b_{\nu,\bf q}^\dagger b_{\mu,\bf k}^\dagger  b^\dagger_{\eta,\bf p} + \textrm{h.c.},
\label{Hss}
\end{equation}
with the  vertex
\begin{equation}
V^{\nu\mu\eta}_{\bf qkp}= -J\sqrt{\frac{3S}{2}}\;
\widetilde{V}^{\nu\mu\eta}_{\bf qkp} \,,
\label{V30}
\end{equation}
and the fully symmetrized dimensionless vertex given by
\begin{eqnarray}
&&\widetilde{V}^{\nu\mu\eta}_{\bf qkp}  =
F^{\nu\mu\eta}_{\bf qkp} (u_{\nu\bf q}+v_{\nu\bf q})(u_{\mu\bf k}v_{\eta\bf p}+v_{\mu\bf k}u_{\eta\bf p})
\nonumber\\
&& \phantom{\widetilde{V}^{\nu\mu\eta}_{\bf qkp}} +
F^{\mu\eta\nu}_{\bf kpq} (u_{\mu\bf k}+v_{\mu\bf k})(u_{\nu\bf p}v_{\eta\bf q}+v_{\nu\bf p}u_{\eta\bf q})
\label{V3}\\
&& \phantom{\widetilde{V}^{\nu\mu\eta}_{\bf qkp}} +
F^{\eta\nu\mu}_{\bf pqk} (u_{\eta\bf p}+v_{\eta\bf p})(u_{\nu\bf q}v_{\mu\bf k}+v_{\nu\bf q}u_{\mu\bf k}) \,,
\nonumber
\end{eqnarray}
for deriving which we have used the symmetry property $F^{\nu\mu\eta}_{\bf qkp}=F^{\nu\eta\mu}_{\bf qpk}$.

Repeating the same  calculation for the $\sqrt{3}\times\sqrt{3}$ state results in the identical expression
of the cubic spin-wave Hamiltonian (\ref{Hss}) and corresponding vertices (\ref{V30}) and (\ref{V3}), but
with  different amplitude $F^{\nu\mu\eta}_{\bf qkp}$
\begin{equation}
F^{\nu\mu\eta}_{\bf qkp} = i\sum_{\alpha\beta}
\epsilon^{\alpha\beta\gamma}\sin(q_{\beta\alpha}) \,
w_{\nu,\alpha}({\bf q}) w_{\mu,\beta}({\bf k}) w_{\eta,\beta}({\bf p})\,.
\label{F3}
\end{equation}
The second-order correction to the ground state energy due to cubic terms is given by
\begin{eqnarray}
\delta E^{(3)}=-\frac{1}{6N}\sum_{\nu\mu\eta}\sum_{{\bf q},{\bf k}}
\frac{|V^{\nu\mu\eta}_{{\bf q},{\bf k},-{\bf k}-{\bf q}}|^2}
{\varepsilon_{\nu,\bf q} + \varepsilon_{\mu,\bf k} + \varepsilon_{\eta,-{\bf k}-{\bf q}}}\, ,
\label{dE3s}
\end{eqnarray}
This energy is per unit cell of 3 spins.
Summation over magnon branches gives 27 individual contributions
of which only 10 are independent.

% ==============================================================================
\subsection{Ordered magnetic moment}
% ==============================================================================

Within the linear SWT, magnetic moment on a site that belongs
to the sublattice $\alpha$ is reduced by zero-point fluctuations
\begin{equation}
\langle S\rangle_\alpha=S-\langle a_{\alpha,i}^\dag a_{\alpha,i}\rangle\,.
\label{Sav}
\end{equation}
Converting from $a_\alpha$ to $d_\mu$ and then to $b_\mu$ operators using unitary (\ref{linearT}) and
then Bogolyubov (\ref{BogolyubovT}) transformations  one arrives to
\begin{equation}
\langle S\rangle_\alpha=S-\frac{1}{N}\sum_{\mu,{\bf k}}w^2_{\mu,\alpha}({\bf k}) \, v^2_{\mu{\bf k}}\,.
\label{Sav1}
\end{equation}
Since all three sublattices are equivalent, symmetrization of (\ref{Sav1}) gives
\begin{equation}
\langle S\rangle=S-\frac{1}{3N}\sum_{\mu,{\bf k}} v^2_{\mu{\bf k}}\,,
\label{Sav2}
\end{equation}
with $v^2_{\mu{\bf k}}$ from (\ref{Bogolyubov_uv}). Since Bogolyubov parameters are implicit functions of
anisotropy $\Delta$, calculations of the magnetization $M\!=\!\langle S\rangle/S$ and the $\langle S\rangle\!=\!0$
Ne\'{e}l order boundary in the $S\!-\!\Delta$ plane can be performed taking the 2D integrals in (\ref{Sav2})  numerically
for the range of $0\!<\!\Delta\!<\!1$.

% ------------------------------------------------------------------------------
\begin{figure}[b]
%\centerline{
\includegraphics[width=0.99\columnwidth]{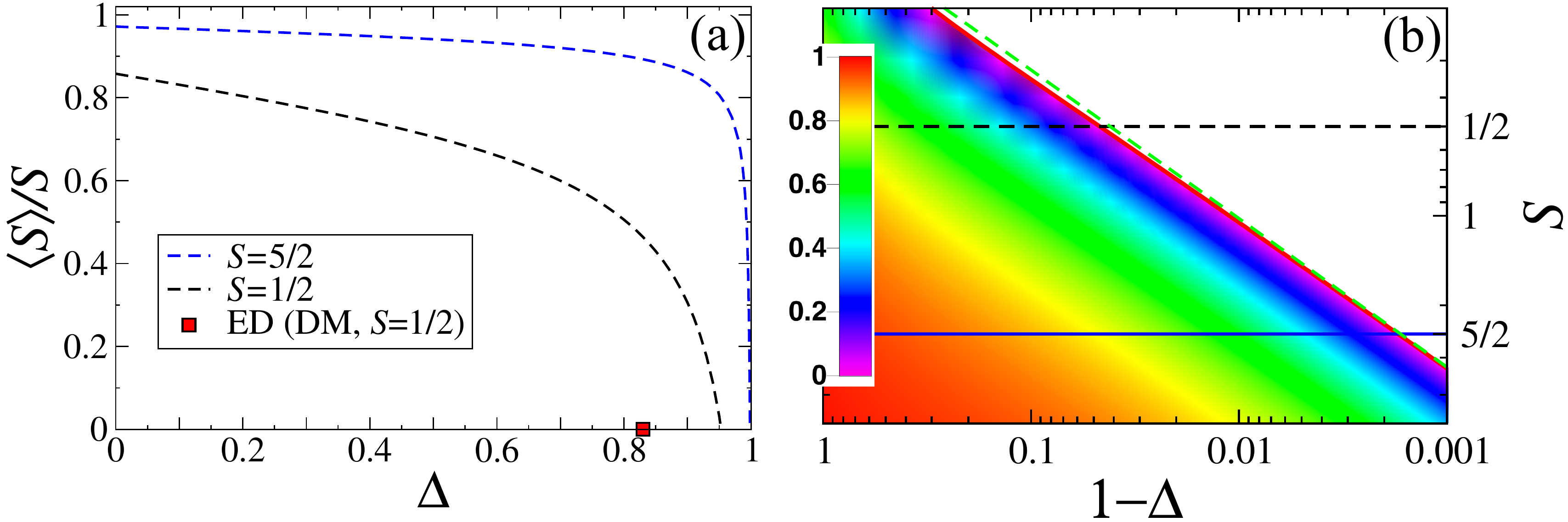}
%}
\caption{(a) Linear SWT result for the magnetization
$\langle S\rangle/S$ vs $\Delta$  for $S\!=\!1/2$ and $S\!=\!5/2$, dashed lines.
Square is  ED result for the case of $S\!=\!1/2$ with DM interaction.
(b) Solid line is the $M\!=\!0$ Ne\'{e}l order phase boundary in
the $S\!-\!\Delta$ plane on the log-log scale, dashed line
is the asymptotic approximation for it; color is for the magnitude of $M$.
}
\label{deltaS}
\end{figure}
% ------------------------------------------------------------------------------

Quantum suppression of the ordered moment  vs anisotropy $\Delta$ is shown in Fig.~\ref{deltaS}(a)   for two values of spin.
Quantum correction diverges for $\Delta \to 1$ due to vanishing energy
of the ``flat mode,'' suggesting a disordered state near the Heisenberg limit for all spins.
The critical value $1-\Delta_c \approx 0.047$ for $S\!=\!1/2$ is also compared with the result for
the Dzyaloshinskii-Moriya (DM) coupling $D_c =0.1$ (with the proper  rescaling
$1\!-\!\Delta\Leftrightarrow\sqrt{3}D$),
found by Exact Diagonalization (ED)  \cite{Cepas08s,Messio10s}.

Near the Heisenberg limit one can neglect non-divergent terms in the integrals in (\ref{Sav2}) and find an
asymptotic expression for the Ne\'{e}l order boundary from
\begin{equation}
\langle S\rangle=0\approx S-\frac{1}{6}\frac{A_1}{\omega_1} \,,
\label{Sav3}
\end{equation}
where $A_1=3/4$, see (\ref{AB1}), and $\omega_1=\sqrt{3(1-\Delta)/2}$, see (\ref{w1}), leading to
\begin{equation}
1-\Delta_c\approx\frac{1}{96S^2} \,,
\label{Sav4}
\end{equation}
which is shown in Fig.~\ref{deltaS}(b) together with the result of the numerical integration in (\ref{Sav2}).

In the same spirit, we attempt to capture the non-perturbative effects of order-by-disorder
mechanism at the Heisenberg limit by using the self-consistently renormalized spin-wave dispersion
of the ``flat mode'' from \cite{Chubukov92s} instead of the dispersionless mode in (\ref{w1})
for the calculation of the mean-field Ne\'{e}l order boundary $\langle S\rangle\!=\!0$ from (\ref{Sav2}).
This procedure should provide a reasonable estimate on the extent of the region of stability due to quantum order-by-disorder
for $\Delta\!=\!1$. In Ref.~\cite{Chubukov92s}, the renormalized dispersion of the ``flat mode'' is
\begin{equation}
\widetilde{\varepsilon}_{1,\bf k}  = 2JS^{2/3}C_{SW}\sqrt{2}\,\omega_{2,\bf k} \ ,
\end{equation}
where $\omega_{2,\bf k}=\sqrt{1-\gamma_{\bf k}}$  and the constant $C_{SW}$ is defined numerically
for two ordered phases: $C_{SW}^{\sqrt{3}\times\sqrt{3}}\!\approx\!0.40$ and
 $C_{SW}^{{\bf q}=0}\!\approx\!0.42$.

Again, neglecting the non-divergent terms in (\ref{Sav2}) reduces $\langle S\rangle=0$ to
\begin{equation}
\langle S\rangle=0\approx S-\frac{1}{6N}\sum_{{\bf k}} \frac{2JSA_1}{\widetilde{\varepsilon}_{1,\bf k}}\,,
\label{Sav_sc}
\end{equation}
which gives
\begin{equation}
S_c\approx \left(\frac{c_0}{8\sqrt{2}\, C_{SW}}\right)^{3/2}\,,
\label{Sc_sc}
\end{equation}
where $c_0=\sum_{\bf k}1/\sqrt{1-\gamma_{\bf k}}=1.4296$.
The resulting values for the
``critical'' $S_c$ come out as  $S_c^{{\bf q}=0}\!\approx\!0.17$
and $S_c^{\sqrt{3}\times\sqrt{3}}\!\approx\!0.18$. Observe that
$S_c^{{\bf q}=0}<S_c^{\sqrt{3}\times\sqrt{3}}$.
While SWT approach clearly exaggerates the extent of the ordered phase,
this estimate makes it extremely unlikely that the Heisenberg KGAFM with $S\!\agt\!1$ will be
magnetically disordered.

% ==============================================================================
\section{Real-space perturbation theory for kagom\'e antiferromagnet}
% ==============================================================================

We develop a real-space perturbation expansion around the manifold of  classical ground
states \cite{Long89s,Heinila93s,Canals04s,Bergman07s}
to determine the ordering pattern in the $XXZ$ kagom\'e-lattice antiferromagnet.
By analyzing perturbative terms of various order we suggest
a real-space hierarchy of effective couplings that are responsible
for the groundstate selection and give a simple qualitative
explanation  of the stability of the ${\bf q}=0$ state for $\Delta=1/2$.

Since the easy-plane anisotropy $\Delta < 1$ confines spins to $xy$ plane,
we consider only planar spin configurations. Geometry of the kagom\'e
lattice allows decomposition of the nearest-neighbor spin Hamiltonian (\ref{Has})
into the sum over triangles, see, e.g., \cite{Chalker92s}, such that the
classical Hamiltonian for an arbitrary planar state with ${\bf S}_i = (S_i^x,S_i^y,0)$
can be written as
\begin{equation}
\hat{\cal H}_{\rm cl} = \frac{J}{2} \sum_\triangle {\bf S}_\triangle^2 - J \sum_i {\bf S}_i^2 \, .
\label{Ecl}
\end{equation}	
Classical energy is minimized for ${\bf S}_\triangle = 0$, whereas the second term in (\ref{Ecl})
gives a constant $E_{\rm cl}/N =-JS^2$. The constraint ${\bf S}_\triangle = 0$
forces spins on every triangle to form a 120$^\circ$ structure, but
the translational pattern remains undetermined leading to a macroscopic number of degenerate ground states
$W \approx 1.13471^N$ [$(\ln W)/N = 0.126377...$]
\cite{Baxter70s,Huse92s}. Degneracy at the level of two triangles sharing a vertex is illustrated in
Figs.~\ref{fig:kagdeg}(a) and (b) and correspond to $\sqrt{3}\times\sqrt{3}$ and ${\bf q}=0$ states, respectively.

The local  field acting on an individual spin is
\begin{equation}
{\bf h}_i = -\frac{\partial E_{\rm g.s.}}{\partial {\bf S}_i} = 2JS {\bf n}_i \quad
{\rm with} \quad
{\bf n}_i = {\bf S}_i/|{\bf S}_i|\ .
\end{equation}
Because of the classical constraint only the last term in  (\ref{Ecl})
contributes to $h\!=\!|{\bf h}_i|\!=\!2JS$. The mean-field theory neglects fluctuations of $h$
and does not lift the degeneracy, hence the need to include correlated fluctuations of spins.
To construct perturbative expansion we use the Hamiltonian in the rotating local frame (\ref{Hs})
and rearrange terms utilizing  the value of the local field $h=2JS$
\begin{eqnarray}
&& \hat{\cal H} =  h \sum_i \delta S_i^z +
J \sum_{\langle ij\rangle}  \Bigl( \Delta S_i^yS_j^y + S_i^xS_j^x\cos\theta_{ij} \nonumber\\
&&\phantom{\hat{\cal H} =}+ \delta S_i^z \delta S_j^z\cos\theta_{ij} +
\sin\theta_{ij} \left(S_i^zS_j^x-S_i^xS_j^z\right)\Bigr)\,,\ \ \ \ \ \ \ \
\label{local1}
\end{eqnarray}
where the classical energy $E_{\rm cl}$ is neglected and we introduce $\delta S_i^z\!=\!S\!-\!S_i^z$.
Note that for any 120$^\circ$ state
$\cos\theta_{ij} \equiv -1/2$ whereas $\sin\theta_{ij} =\pm \sqrt{3}/2$.

The first term in (\ref{local1}) includes only on-site spin fluctuations
and is chosen as the unperturbed Hamiltonian
\begin{equation}
\hat{\cal H}_0 = h \sum_i \delta S_i^z \  ,
\label{H01}
\end{equation}
while the  perturbation is naturally divided into four parts
$\hat{V} \!=\!  \sum_{i,j}\left(\hat{V}^{ij}_{1}\!+\!\hat{V}^{ij}_{2}\!+\!\hat{V}^{ij}_{3}\!+\! \hat{V}^{ij}_{4}\right)$
with
\begin{eqnarray}
%&&\hat{V} =  \sum_{i,j}\left(\hat{V}^{ij}_{1}+\hat{V}^{ij}_{2}+\hat{V}^{ij}_{3}+ \hat{V}^{ij}_{4}\right)
%\, , \nonumber\\
&&\hat{V}^{ij}_{1} =  -\frac{J}{8}\Bigl(\Delta+\frac{1}{2}\Bigr)
\bigl(S_i^+S_j^+ + \mbox{H.c.} \bigr)\, , \nonumber\\
&&
\hat{V}^{ij}_{2}  =  \frac{J}{4}\Bigl(\Delta-\frac{1}{2}\Bigr) S_i^+S_j^-\,,
\label{V} \\
&& \hat{V}^{ij}_{3} = -\frac{J}{2}\sin\theta_{ij}\, \delta S_i^z\,\bigl(S_j^++S_j^-\bigr),\ \
\hat{V}^{ij}_{4}  =  -\frac{J}{2} \delta S_i^z\, \delta S_j^z\,, \nonumber
\end{eqnarray}
where we have used $\cos\theta_{ij}=-1/2$ and left explicit $\sin\theta_{ij}=\pm\sqrt{3}/2$ in
$\hat{V}_3$. Since the ordering patterns differ by clockwise or counter-clockwise rotation of spins, they can be
distinguished only by the $\hat{V}_{3}$ operator, which, therefore,  plays the key role in the groundstate selection.
The first three  terms in (\ref{V}) can be referred to as the  double spin-flip,
spin-flip hopping, and  single-flip, the latter is a descendant of the cubic term in (\ref{Hs}).

The energy corrections generated by the expansion in $\hat{V}$ can be obtained from the standard
perturbation theory.
One can formulate several simple rules, which help to identify contributions that are relevant
to the groundstate selection: \\[1.0mm]
1) Every member of the perturbation series must be represented by a linked cluster, ensuring
that energy is extensive, $\delta E \sim N$, with each link corresponding to the action of one of
the perturbation terms,  $\hat{V}_{n}^{ij}$, acting on a specific lattice bond $(i,j)$.
The total number of links is equal to the order of  expansion.
Several links on the same bond  are allowed. \\[1mm]
2) Any groundstate $|0\rangle$ from the classical 120$^\circ$ manifold  is a vacuum for spin flips because
all spins are fully oriented along their local fields, hence $\langle 0|\hat{V}|0\rangle=0$.
A simple inspection of $\hat{V}$ in (\ref{V}) shows that
any term in the expansion must begin and end with the double spin-flip
$\hat{V}_{1}$. \\[1mm]
3) Since the single-flip term $\hat{V}_{3}$ is the only one odd in the number of spin flips,
any relevant term of the expansion must contain an even number of them. \\[1mm]
4) We are looking for the lowest-order energy correction  which lifts the
degeneracy between classical ground states. The correction
of the $p$th order associated with a specific linked cluster is given by
\begin{equation}
\delta E^{(p)} = \sum_{n_k}\frac{\langle 0| \hat{V}|n_1\rangle\langle n_1| \hat{V}|n_2\rangle\ldots
\langle n_{p-1}| \hat{V}|0\rangle}{(E_0-E_{n_1})\ldots(E_0-E_{n_{p-1}})} \,.
\label{dE}
\end{equation}
Here, $E_0$ is the classical ground-state energy and $E_{n_k}$ are the unperturbed energies of
excited states. This expression is straightforwardly obtained from
the Brillouin-Wigner theory by replacing the exact ground-state energy
$E$ with $E_0$, which is justified because $E_{n_k}$ are the same for all classical ground states.

% ==============================================================================
\subsection{Degeneracy lifting}
% ==============================================================================

% -----------------------------------------------------------------------------
\begin{figure}[t]
%\centerline{
\includegraphics[width=0.99\columnwidth]{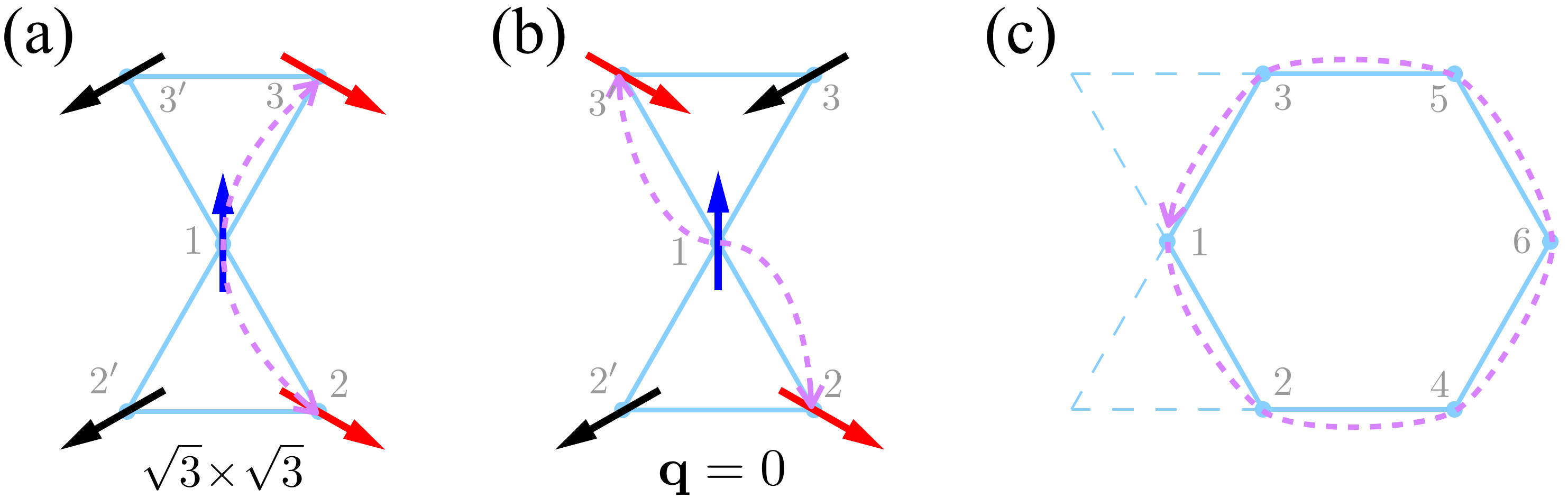}
%}
\caption{
Two degenerate spin configurations of the kagom\'e-lattice antiferromagnet with
(a) ferromagnetic-like  ($\sqrt{3}\times\!\sqrt{3}$) and (b) antiferromagnetic-like (${\bf q}=0$)
arrangement of second-neighbor spins. (c)
Smallest cluster with nontrivial topology contributing to the energy difference
between states (a) and (b) in the 7th order of expansion.
Schematics of the tunneling processes are also shown.
}
\label{fig:kagdeg}
\end{figure}
% ------------------------------------------------------------------------------

The degeneracy of the ordering patterns in the kagom\'e-lattice antiferromagnet is illustrated in
Figs.~\ref{fig:kagdeg}(a) and (b).
For a fixed spin triad  in the lower triangle, spin ${\bf S}_2$ can be parallel either
to spin ${\bf S}_3$  or ${\bf S}_{3'}$ of the upper triangle, having ferromagnetic-like or antiferromagnetic-like
alignment of the second-neighbor spins. Extended over the entire lattice, the two structures correspond
to the $\sqrt{3}\times\!\sqrt{3}$ and ${\bf q}=0$ states.

According to the rules formulated above, the lowest-order  correction distinguishing
between the two patterns may appear only in the fourth order.
An example of such a tunneling process is given by the operator sequence acting on the ground state from left to right
\begin{equation}
\hat{V}_{1}^{12} \to  \hat{V}_{3}^{13} \to \hat{V}_{3}^{12}\to  \hat{V}_{1}^{13} \, ,
\end{equation}
see Fig.~\ref{fig:kagdeg}(a).
The respective energy shift depends explicitly on the mutual orientation of ${\bf S}_2$
and ${\bf S}_3$ because
$\delta E^{(4)} \!\propto \! \sin \theta_{12} \sin\theta_{13}$.
However, an obvious symmetry leaves the degeneracy intact at this order of expansion,
because for any coupling between ${\bf S}_2$ and ${\bf S}_3$ there is a
``mirror'' counterpart to the tunneling process that couples  ${\bf S}_2$ with ${\bf S}_{3^\prime}$ identically,
see Fig.~\ref{fig:kagdeg}(b), yielding $\delta E^{(4)} \!\propto \! \sin \theta_{12} \sin\theta_{13'}$.
Since for any classical ground state $\sin\theta_{13'}=-\sin\theta_{13}$, the two processes provide
the same energy gain to both
$\sqrt{3}\times\!\sqrt{3}$ and ${\bf q}\!=\!0$ states.

Generalizing this trend to the higher-order terms having the  form  $\delta E^{(p)}\! \sim \!
\sin \theta_{12} \sin\theta_{13}$, we conclude that
the processes represented by the graphs with trivial
topology, i.e. connecting sites 2 and 3 via site 1 only, have to be discarded because there always exists a
mirror graphs that connect site 2 and 3' via precisely the same process.

Therefore, the tunneling paths which lift the degeneracy between
$\sqrt{3}\times\!\sqrt{3}$ and ${\bf q}\!=\!0$ states must
have a non-trivial topology, with the shortest one making a loop around a hexagon, see Fig.~\ref{fig:kagdeg}(c).
Because  such processes also need to be proportional to $\sin \theta_{12} \sin\theta_{13}$, they appear
in the seventh order of expansion with one of the hexagon sides
containing a double link.  A simple analysis shows that the double link must be located
on one of the two bonds: (1,2) or (1,3).
According to the rules formulated above, all relevant
seventh-order perturbation terms have to contain two double-flips $\hat{V}_{1}$ and
two single-flips $\hat{V}_{3}$, the latter acting only on bonds (1,2) and (1,3). Then, the remaining
three links must contain either double spin-flip
$\hat{V}_{1}$ or spin-flip hopping $\hat{V}_{2}$---the fourth term in (\ref{V}), $\hat{V}_{4}$, does
not contribute to this order in $1/S$.

Note that there are close parallels between the non-linear SWT and the real-space approach.
Although the degeneracy-lifting contribution in the latter is of seventh order, it is still
of second order in the cubic terms $\hat{V}_{3}$,  same as in the non-linear SWT (\ref{dE3s}).

The number of double-flips $\hat{V}_{1}$ or spin-flip hoppings $\hat{V}_{2}$ in a specific
tunneling process determines the sign of the corresponding energy correction (\ref{dE}).
In fact, according to (\ref{V}), every double-flip operator carries a minus
sign, whereas spin-flip hopping has a prefactor $(\Delta-\frac{1}{2})$, which is
positive or negative depending on the value of $\Delta$. For the seventh-order processes, denominator in (\ref{dE})
contains the product of six negative factors  and is
positive.

% -----------------------------------------------------------------------------
\begin{figure}[t]
%\centerline{
\includegraphics[width=0.99\columnwidth]{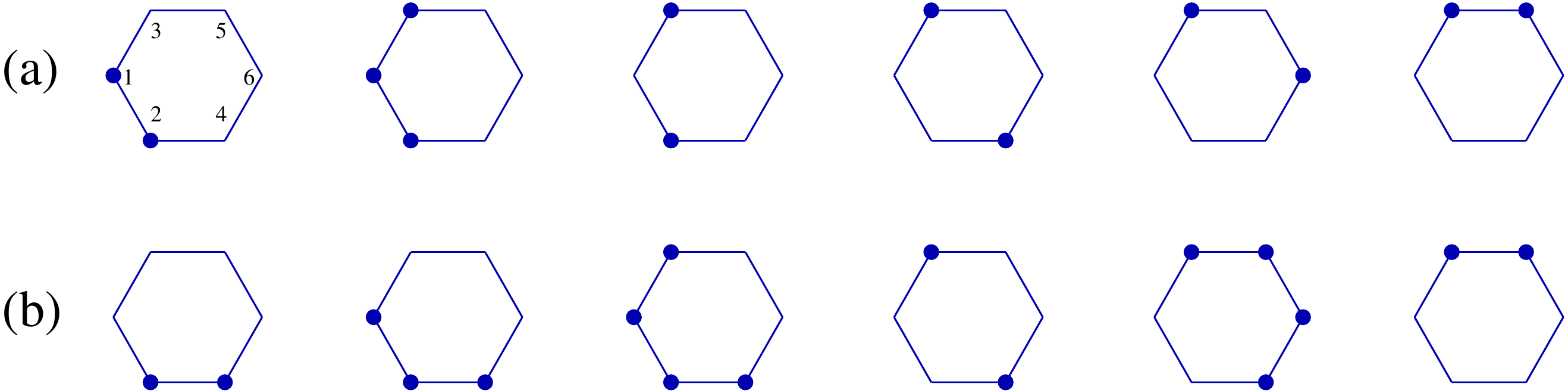}
%}
\caption{
Intermediate excited states of the 7th order tunneling process involving hexagon cluster
generated by different
operator sequences: (a) with three spin-flip hoppings, (b)  without hoppings, all spin-flips. 
Position of spin flips  is indicated by filled circles.
}
\label{fig:interm}
\end{figure}
% ------------------------------------------------------------------------------

One example of the possible 7th order processes, which involves two
double spin flips, two single flips, and three spin-flip hops,
is given by the operator sequence (from left to right)
\begin{equation}
\hat{V}_{1}^{12} \to  \hat{V}_{3}^{13} \to \hat{V}_{3}^{21}\to
\hat{V}_{2}^{24} \to  \hat{V}_{2}^{46} \to \hat{V}_{2}^{65}\to\hat{V}_{1}^{35}.
\label{V7a}
\end{equation}
The intermediate stages of this perturbation process are
shown in Fig.~\ref{fig:interm}(a) where positions of spin flips is indicated by circles.
Using $h=2JS$ and calculating  the matrix elements
we obtain
\begin{equation}
\delta E^{(7)}_a = - \frac{J}{2^{17}\cdot 3}\,
(\Delta +{\textstyle \frac{1}{2}})^2(\Delta -{\textstyle \frac{1}{2}})^3
 \sin \theta_{12}\sin \theta_{13}.
\label{dE7a}
\end{equation}
One can also easily determine the respective multiplicative factor $2^5$ associated with
permutations of operators in (\ref{V7a}),
though our subsequent conclusions do not depend on precise numbers.
For $\Delta>1/2$, the energy correction (\ref{dE7a}) is negative
and favors ferromagnetic alignment of second-neighbor spins, see Fig.~\ref{fig:kagdeg}(a),
i.e., the $\sqrt{3}\times\!\sqrt{3}$ structure. For $\Delta<1/2$, the energy correction
changes sign and  favors the ${\bf q}\!=\!0$ state of  Fig.~\ref{fig:kagdeg}(b).

Clearly, $\Delta=1/2$ is special, because the spin-flip
hopping amplitude $V_2^{ij}$ vanishes.
At this value of anisotropy, the non-vanishing 7th order processes  must involve 
only double spin flips and two single spin-flips. An example of such process is given by
\begin{equation}
\hat{V}_{1}^{24} \to  \hat{V}_{3}^{21} \to \hat{V}_{3}^{13}\to
\hat{V}_{1}^{12} \to  \hat{V}_{1}^{56} \to \hat{V}_{1}^{46}\to\hat{V}_{1}^{35},
\label{V7b}
\end{equation}
with intermediate states shown in Fig.~\ref{fig:interm}(b) and the energy shift
\begin{equation}
\delta E^{(7)}_b = \frac{J}{2^{19}\cdot 3}\,
(\Delta +{\textstyle \frac{1}{2}})^5
 \sin \theta_{12}\sin \theta_{13}.
\label{dE7b}
\end{equation}
Straightforward but tedious calculation of the  multiplicative factor 
attributed to (\ref{dE7b})
yields $176/3$.
Because of the positive sign, the correction $\delta E^{(7)}_b$ corresponds to the antiferromagnetic 
effective interaction
between second-neighbor spins, Fig.~\ref{fig:kagdeg}(b), for the entire range $0<\Delta<1$.
Hence,  we can claim that for $\Delta=1/2$
quantum fluctuations stabilize the ${\bf q}\!=\!0$ structure.

Furthermore,  for $0\!<\!\Delta\!<\!1/2$, the spin-flip hopping amplitude $V_2^{ij}$ has 
the same negative sign as $V_1^{ij}$, making the sign of 
all 7th-order perturbation processes contributing to the quantum energy 
shift the same, thus  favoring the ${\bf q}\!=\!0$ state.
We emphasize again that this conclusion relies only on the sign of the matrix elements
and the order of the perturbation
process, which, in turn, depends on the length of the shortest topologically nontrivial 
loop in the lattice. Thus, the quantum selection of the ${\bf q}\!=\!0$ state for
$0\!<\!\Delta\!<\!1/2$ stems from the lattice geometry. 

For $\Delta>1/2$, the  perturbation terms with odd number of spin-flip
hops change sign and favor  the
$\sqrt{3}\!\times\!\sqrt{3}$ state.
This implies that the transition between the two magnetic structures can
only happen at $\Delta_c\!>\!1/2$ whose value must be determined
by summing all contributions including corresponding multiplicative factors.
This is, again, in accord with the answer from the second-order non-linear SWT, $\Delta_c\!\approx\!0.72$.

Another close parallel between the non-linear SWT and the real-space approach is worth noting.
The high order of the  tunneling processes relevant to the degeneracy lifting explains the origin of the smallness
of the quantum order-by-disorder effect. Since the real-space perturbation theory is, essentially, an expansion in
coordination number ($z\!=\!4$ for the kagom\'{e} lattice), a rough but intuitively straightforward
estimate of the seventh-order process, taking into account the number of next-nearest neighbor bonds 
and multiplicative factor of the symmetry-related processes, 
gives $\delta E\!=\!8J/z^7\!\approx\!5\cdot 10^{-4}J$ per spin. This is
in a very good agreement with the results of SWT, $\approx\! 5\cdot 10^{-4}J$ for $\Delta=1/2$.
Obviously, such a close quantitative agreement with the naive estimate is simply fortuitous. 
A more careful calculation for $\Delta=1/2$, using our results in (\ref{dE7b}) with the 
combinatorial factor gives $\delta E\!\approx\!1.1\cdot 10^{-4}J$. This is in a good qualitative agreement 
with the SWT results and also implies that the higher-order ``dressings'' of the loop-like processes are quantitatively 
important.

A couple of additional remarks concern the ``third-neighbor'' effective interaction of
spins 2 and 3', see Fig.~\ref{fig:kagdeg}. The corresponding linked cluster has one more link
and hence appears only in the eighth order of perturbative expansion.
Moreover, because of the even number of sites, all  8th-order
processes must involve at least one   spin-flip hopping operator $\hat{V}_{2}$.
Thus, for $\Delta=1/2$ the 8th-order energy shift
$\delta E^{(8)} \sim \sin \theta_{12}\sin \theta_{13'}$ vanishes and one has to
go to the 9th order and the larger cluster. All that indicates
that the ${\bf q}=0$ spin structure should be stable with respect to the higher-order corrections.

% ==============================================================================

% ==============================================================================

\begin{thebibliography}{99}
\bibitem{footnote_Dz}
We note that there exist a close similarity of the harmonic part of the $XXZ$ model and the
kagom\'{e} Hamiltonian  with the out-of-plane DM interaction $D_z$. However,
in the latter case the ground state  is selected on a classical level.
In that sense, $XXZ$ anisotropy is a much lighter perturbation
as it does not lift the macroscopic degeneracy between the manifold of $120^\circ$ states
and allows to study a purely quantum ObD mechanism.
\bibitem{footnote_1} Thus, for the values of $\Delta<\Delta_c\approx 0.72$, 
one can expect a transition, as a function of $T$, between the quantum-fluctuation-dominated ${\bf q}=0$ correlations 
to thermal-fluctuation-dominated $\sqrt{3}\times\sqrt{3}$ correlations. While the transition is likely to be of the first order,  
this is  a matter for a  future investigation.
\end{thebibliography}
\end{document}